\documentclass[12pt]{article}

\usepackage{amssymb}
\usepackage{amsthm}
\newtheorem {theorem} {Theorem}

\newtheorem {proposition} [theorem]{Proposition}

\newtheorem {remark} [theorem]{\bf Remark}

\title{Perturbed rank 2 Poisson systems and periodic orbits on Casimir invariant manifolds}

\author{Isaac A. Garc\'{\i}a$^{\ (1,*)}$ \hspace{5mm}  Benito Hern\'andez-Bermejo$^{\ (2)}$}

\date{{\small $^{\ (1)}$ Departament de Matem\`atica. Universitat de
Lleida. \\ Avda. Jaume II, 69. 25001 Lleida, Spain.
\\ E--mail: {\tt garcia@matematica.udl.cat}
\\ $ $ \\
$^{\ (2)}$ Departamento de Biolog\'{\i}a y Geolog\'{\i}a, F\'{\i}sica y Qu\'{\i}mica
Inorg\'{a}nica.
\\ Universidad Rey Juan Carlos.
\\ Calle Tulip\'{a}n S/N. 28933--M\'{o}stoles--Madrid, Spain.
\\ E-mail: {\tt benito.hernandez@urjc.es}}}

\begin{document}

\maketitle

\begin{abstract}
A class of $n$-dimensional Poisson systems reducible to an unperturbed harmonic oscillator shall be considered. In such case, perturbations leaving invariant a given symplectic leaf shall be investigated. Our purpose will be to analyze the bifurcation phenomena of periodic orbits as a result of these perturbations in the period annulus associated to the unperturbed harmonic oscillator. This is accomplished via the averaging theory up to an arbitrary order in the perturbation parameter $\varepsilon$. In that theory we shall also use both branching theory and singularity theory of smooth maps to analyze the bifurcation phenomena at points where the implicit function theorem is not applicable. When the perturbation is given by a polynomial family, the associated Melnikov functions are polynomial and tools of computational algebra based on Gr\"obner basis are employed in order to reduce the generators of some polynomial ideals needed to analyze the bifurcation problem. When the most general perturbation of the harmonic oscillator by a quadratic perturbation field is considered, the complete bifurcation diagram (except at a high codimension subset) in the parameter space is obtained.
Examples are given.
\end{abstract}


\mbox{}



\vfill

\footnoterule

$^*$ Corresponding author. Telephone: (+34) 973702728. Fax: (+34) 973702702.

\pagebreak

\section{Introduction}

Finite-dimensional Poisson systems (see \cite{olv1,wei1} and references therein for an over\-view) have a significant presence in most domains of physics and applied mathematics. The specific format of Poisson systems has allowed the development of many tools for their analysis (for instance, see
\cite{iyb1}-\cite{iyb3},\cite{bs1}-\cite{bs4},\cite{olv1} and references therein for a sample). In addition, the relevance of Poisson systems arises from the fact that they constitute a generalization of classical Hamiltonian systems comprising nonconstant structure matrices as well as odd-dimensional vector fields. Additionally, Poisson system format is invariant under general diffeomorphic transformations, therefore not being restricted to the use of canonical transformations.

Consider a smooth vector field having a finite-dimensional Poisson structure
\begin{equation}
\label{poisson-V-1}
	\frac{\mbox{\rm d}x}{\mbox{\rm d}t} = {\cal J}(x) \cdot \nabla H (x)
\end{equation}
of dimension $n$ and rank $r = 2 \leq n$ constant in an open set $\Omega \subseteq \mathbb{R}^n$. In (\ref{poisson-V-1}) ${\cal J}(x)$ and $H(x)$ are the structure matrix and Hamiltonian function, respectively. Then under these hypotheses for each point $x_0 \in \Omega$ there is (at least locally in a neighborhood $\Omega_0 \subset \Omega$ of $x_0$) a complete set of functionally independent Casimir invariants $\{ D_{3}(x), \ldots , D_n(x) \}$ in $\Omega_0$, as well as a transformation $x \mapsto \Phi_D(x) = y$ where $\Phi_D$ is a smooth diffeomorphism in $\Omega_0$ bringing the system (\ref{poisson-V-1}) into its Darboux canonical form. Thus, beyond the fact that Poisson systems are a formal generalization of classical Hamiltonian flows, Darboux Theorem provides the dynamical basis for such a generalization.

In this article, a class of $n$-dimensional Poisson systems reducible to an unperturbed harmonic oscillator shall be considered. In such case, perturbations leaving invariant a given symplectic leaf shall be investigated. Our purpose will be to analyze the bifurcation phenomena of periodic orbits as a result of these perturbations in the period annulus associated to the unperturbed harmonic oscillator. This is accomplished via the averaging theory up to an arbitrary order in the perturbation parameter $\varepsilon$. In that theory we shall also use both branching theory and singularity theory of smooth maps to analyze the bifurcation phenomena at points where the implicit function theorem is not applicable. When the perturbation is given by a polynomial family, the associated Melnikov functions are polynomial and tools of computational algebra based on Gr\"obner basis are employed in order to reduce the generators of some polynomial ideals needed to analyze the bifurcation problem. When the most general perturbation of the harmonic oscillator by a quadratic perturbation field is considered, the complete bifurcation diagram (except at a high codimension subset) in the parameter space is obtained.

\section{Reduction procedure}
\subsection{Darboux canonical form and harmonic oscillator form}

For the Poisson system (\ref{poisson-V-1}) and under the assumptions previously stated, we shall say that the Hamiltonian function $H(x)$ is {\it quasi-harmonic} for variables $x_i$ and $x_j$ in a domain $\Omega \subset \mathbb{R}^n$ if, by definition, it can be written in the form: $H(x) \equiv H(\varphi(x_i, x_j), D_3(x),\ldots, D_n(x))$ where $\varphi(x_i, x_j)$ admits, in the region of interest $\Omega$, at least one decomposition of the kind $\varphi(x_i, x_j) = \varphi_1(x_i, x_j) + \varphi_2(x_i, x_j)$ with $\varphi_k(x_i, x_j) > 0$ for $k=1,2$, and with application $(x_i, x_j) \mapsto (\varphi_1(x_i, x_j), \varphi_2(x_i, x_j))$ being invertible.

Let us then assume, without loss of generality, a Hamiltonian quasi-harmonic for $x_1$ and $x_2$. The following change of variables is to be performed:
\begin{equation}
\label{Change-FCD}
x \mapsto y = (y_1, \ldots, y_n) = \Phi(x) = \left( \sqrt{2 \varphi_1(x_1, x_2)}, \sqrt{2 \varphi_2(x_1, x_2)},  D_3(x),\ldots, D_n(x) \right)
\end{equation}
By definition, $\Omega$ is the open set such that the Poisson system (\ref{poisson-V-1}) is defined and has rank $r=2$, and in addition $\left. \Phi \right| _\Omega$ is a diffeomorphism. In particular, $\Phi$ is a diffeomorphism in $\Omega$ under the hypotesis that the Jacobian verifies
$$
\det \left(\frac{\partial(D_3, \ldots, D_n)}{\partial(x_3, \ldots, x_n)} \right) \neq 0 \ \ \mbox{in} \ \Omega.
$$
Then the transformed system can be written as
\[
	\frac{\mbox{\rm d}y}{\mbox{\rm d}t} = {\cal J}^*(y) \cdot \nabla H^*(y)
\]
with $H^*(y)=H \circ \Phi^{-1}(y) = \hat{H}\left(\frac{1}{2}(y_1^2+y_2^2), y_3, \ldots, y_n \right)$, and ${\cal J}^*(y)= \eta (y) \cdot {\cal J}_D$, where
\[
	{\cal J}_D \equiv
	\left( \begin{array}{cc} 0 & 1 \\ -1 & 0 \end{array} \right)
	\oplus {\cal O}_{n-2} =
\left( \begin{array}{cccc}
      0  & 1 & \vline & \mbox{} \\
      -1 & 0 & \vline & \mbox{} \\ \hline
      \mbox{} & \mbox{} & \vline & {\cal O}_{n-2}
      \end{array} \right)
\]
is the Darboux canonical form matrix for the rank-2 case, where ${\cal O}_{n-2}$ denotes the null square matrix of order $n-2$. Finally, rescaling the time as $t \mapsto \tau$ with $d \tau = \eta \; d t$ we obtain the Darboux canonical form in $\Omega$ of the Poisson system (\ref{poisson-V-1}).

Moreover, we can proceed further and reduce the system completely to a classical harmonic oscillator. For this, we first rectrict ourselves to one symplectic leaf $y_i=y_i(0)=c_i$, for $i=3, \ldots ,n$. We are thus left with a planar classical Hamiltonian system for which the structure matrix is the $2 \times 2$ symplectic matrix, and the Hamiltonian is $\tilde{H} \left( \frac{1}{2}(y_1^2+y_2^2) \right) \equiv \hat{H}\left(\frac{1}{2}(y_1^2+y_2^2), c_3, \ldots, c_n \right)$. Now denote as $\tilde{H}'(z) = d \tilde{H}(z)/dz$. Then the reduction is completed \cite{bs2} by means of an additional time reparametrization $\tau \mapsto \rho$ with $d \rho = \mu(y_1,y_2) \; d \tau$, where $\mu(y_1,y_2) = \tilde{H}' \left( \frac{1}{2}(y_1^2+y_2^2) \right)$. The outcome is a one degree of freedom harmonic oscillator of Hamiltonian $\mathcal{H}(y_1, y_2) = \frac{1}{2} (y_1^2+y_2^2)$ and time $\rho$.

\subsection{Perturbations leaving invariant a given simplectic leaf}

We consider now the analytical perturbations of the initial Poisson system (\ref{poisson-V-1})
\begin{equation}
\label{poisson-aver-1}
	\frac{\mbox{\rm d}x}{\mbox{\rm d}t} = {\cal J}(x) \cdot \nabla H (x) + \varepsilon F(x; \varepsilon)
\end{equation}
where $\varepsilon \neq 0$ is a small perturbation real parameter and $F$ is an analytic vector field in $\Omega$ depending analytically on the parameter $\varepsilon$ and satisfying $F(0; \varepsilon) = 0$ and $\nabla_x F(0; \varepsilon) = 0$. Performing the Darboux canonical form reduction of the previous subsection, we obtain that (\ref{poisson-aver-1}) becomes the analytic system
\begin{equation}
\label{poisson-aver-4}
   \frac{\mbox{\rm d}y}{\mbox{\rm d} \tau} = {\cal J}_D \cdot \nabla H^*(y) + \varepsilon F^*(y; \varepsilon)
\end{equation}
defined in $\Omega^* = \Phi(\Omega)$.

Let us choose an invariant simplectic leaf $\mathcal{L}_c = \cap_{j=3}^n \{D_j(x) = c_j\}$ of the Poisson system (\ref{poisson-V-1}) for certain $c =(c_3, \ldots, c_n) \in \mathbb{R}^{n-2}$ such that  $\mathcal{L}_c \cap \Omega \neq \emptyset$. Assume moreover that the perturbation field $F(x; \varepsilon)$ is such that $\mathcal{L}_c$ becomes an invariant surface of the perturbed system (\ref{poisson-aver-1}). Under these conditions, diffeomorphism $\Phi$ defined in (\ref{Change-FCD}) and the rescaling of time $t \mapsto \tau$ previously characterized transform (\ref{poisson-aver-1}) in $\Omega$ into a system in $\Omega^*$ which can be restricted to $\Phi(\mathcal{L}_c)$ leading to a two dimensional system because ${\rm codim}(\mathcal{L}_c) = 2$. More specifically (\ref{poisson-aver-4}) can be written as
\begin{eqnarray}
\frac{dy_1}{d \tau} &=& \frac{\partial H^*}{\partial y_2} + \varepsilon P(y; \varepsilon)  \ , \nonumber \\
\frac{dy_2}{d \tau} &=&  -\frac{\partial H^*}{\partial y_1} + \varepsilon Q(y; \varepsilon)   \ , \nonumber \\
\frac{dy_3}{d \tau} &=& \varepsilon (y_j- c_j) R_j(y; \varepsilon), \ \ j = 3,\ldots, n . \nonumber
\end{eqnarray}
Finally, the restriction to $\Phi(\mathcal{L}_c)$ combined with the time rescaling $\tau \mapsto \rho$ described in the previous subsection leads to
\begin{eqnarray}
\frac{dy_1}{d \rho} &=& \frac{\partial \mathcal{H}}{\partial y_2} + \varepsilon P(y_1, y_2, c; \varepsilon)  \ , \nonumber \\
\frac{dy_2}{d \rho} &=&  -\frac{\partial \mathcal{H}}{\partial y_1} + \varepsilon Q(y_1, y_2, c; \varepsilon)   \ , \label{PlanarSystem*}
\end{eqnarray}
where $\mathcal{H}(y_1, y_2) = \frac{1}{2} (y_1^2+y_2^2)$. The reduction to a perturbed harmonic oscillator is thus accomplished.

\subsection{The Lagrange standard form of averaging theory}

In polar coordinates, $y_1 = r \cos\theta$, $y_2 = r \sin\theta$, system (\ref{PlanarSystem*}) becomes
\begin{eqnarray}
\dot{r} &=& \varepsilon \, G_1^*(\theta, r, c; \varepsilon) \ , \nonumber \\
\dot{\theta} &=& -1 + \frac{\varepsilon}{r} G_2^*(\theta, r, c; \varepsilon)   \ , \label{poisson-aver-2}
\end{eqnarray}
where
\begin{eqnarray*}
G_1^*(\theta, r, c; \varepsilon) &=& \cos\theta \, P(r \cos\theta, r \sin\theta, c; \varepsilon) + \sin\theta \, Q(r \cos\theta, r \sin\theta, c; \varepsilon) \ ,  \\
G_2^*(\theta, r, c; \varepsilon) &=&  \cos\theta \, Q(r \cos\theta, r \sin\theta, c; \varepsilon) - \sin\theta \, P(r \cos\theta, r \sin\theta, c; \varepsilon) \ .
\end{eqnarray*}
Notice that this system is only well defined for $r >0$. Moreover, in this region, since for sufficiently small $\varepsilon$ we have
$\dot{\theta} < 0$ in an arbitrarily large ball centered at the origin, we can
rewrite the differential system (\ref{poisson-aver-2}) in such ball into the form
\begin{equation}\label{poisson-aver-3*}
\frac{d r}{d \theta}  = \varepsilon  \, G(\theta, r, c; \varepsilon)
\end{equation}
by taking $\theta$ as the new independent variable. Recall that any $2 \pi$--periodic solution of (\ref{poisson-aver-3*}) corresponds biunivocally with a periodic orbit of (\ref{poisson-aver-1}) on an arbitrarily large compact set included in $\mathcal{L}_c \cap \Omega$. Therefore, system (\ref{poisson-aver-3*}) is $2 \pi$--periodic in variable $\theta$ and is in the Lagrange standard form of averaging theory.

\subsection{Example: Maxwell-Bloch equations}

The real-valued Maxwell-Bloch system (see \cite{dah} and references therein) is given by the following polynomial vector field in $\mathbb{R}^3$:
\begin{equation}
\label{MB}
\dot{x}_1 = x_2 \:\: , \:\:\: \dot{x}_2 = x_1 x_3 \:\: , \:\:\: \dot{x}_3 = -x_1 x_2.
\end{equation}
Equations (\ref{MB}) can be written as a Poisson system (\ref{poisson-V-1}) with Hamiltonian $H(x) = \frac{1}{2} (x_2^2 + x_3^2)$ and structure matrix
\[
{\cal J}(x) = \left( \begin{array}{ccc} 0 & 1 & 0 \\ - 1 & 0 & x_1 \\0 & - x_1 & 0 \end{array} \right).
\]
Since ${\rm rank}({\cal J}) = 2$ everywhere it has one independent Casimir invariant which can be chosen as
$D(x) = x_3 + \frac{1}{2} x_1^2$.

Let $F(x; \varepsilon) = (A(x), B(x), C(x))$ be the perturbation vector field in (\ref{poisson-aver-1}). We make the following statement: the perturbed field (\ref{poisson-aver-1}) has the invariant surface ${\mathcal L}_c = \{ x \in \mathbb{R}^3 : D(x) = c \}$ for some arbitrary real constant $c \in \mathbb{R}$ if and only if $D(x)-c$ divides the analytic function $x A(x) + C(x)$. The proof is as follows: $S_c$ is an invariant surface of (\ref{poisson-aver-1}) if and only if there is a real analytic function $K$ in $\mathbb{R}^3$ such that $\mathcal{Y}(D(x)-c) = K(x) (D(x)-c)$ where $\mathcal{Y} = A(x) \partial_{x_1} + B(x) \partial_{x_2} + C(x) \partial_{x_3}$ is the vector field (linear differential operator) associated to $F$. Then, direct computations give $x A(x) + C(x) = K(x) (D(x)-c)$ thus proving the claim.

Note that the Maxwell-Bloch Hamiltonian is quasi-harmonic in terms of variables $x_2$ and $x_3$, namely $H(x)=\varphi_1(x_2,x_3)+\varphi_2(x_2,x_3)$, with $\varphi_1(x_2,x_3)= \frac{1}{2}x_2^2$ and $ \varphi_2(x_2,x_3)= \frac{1}{2} x_3^2$. Accordingly,
we can perform the change of variables given by the diffeomorphism $x = (x_1, x_2, x_3) \mapsto y=(y_1, y_2, y_3) = \Phi(x) = (x_2, x_3, D(x))$ defined in the region $\Omega = \{ x \in \mathbb{R}^3 \; : \; x_i \neq 0, \; i=1,2,3 \}$. This is the natural choice in order to arrive to a harmonic oscillator. Observe that under such transformation, the surface $\mathcal{L}_c$ is transformed into the half-plane $\Pi = \{ y \in \Omega^* \subset \mathbb{R}^3 : y_3 = c \}$ defined in $\Omega^* = \Phi(\Omega) = \{ y \in \mathbb{R}^3 : y_1 \neq 0, y_2 \neq  0, y_3 > y_2 \}$. The perturbed system (\ref{poisson-aver-1}) defined in $\Omega^*$ adopts the form
\begin{eqnarray}
\dot{y}_1 &=& \eta(y) \left( \frac{\partial \mathcal{H}}{\partial y_2} + \varepsilon P(y) \right)  \ , \nonumber \\
\dot{y}_2 &=& \eta(y) \left( -\frac{\partial \mathcal{H}}{\partial y_1} + \varepsilon Q(y) \right)  \ , \label{MB1} \\
\dot{y}_3 &=& \varepsilon (y_3- c) R(y), \nonumber
\end{eqnarray}
where $\eta(y) = \sqrt{2(y_3-y_2)} > 0$ in $\Omega^*$, $\mathcal{H}(y_1, y_2) = \frac{1}{2} (y_1^2+y_2^2)$, $P(y) = B(\eta(y), y_1, y_2)$, $Q(y) = - \eta(y) A(\eta(y), y_1, y_2) + (y_3-c) K(\eta(y), y_1, y_2)$ and $R(y)$ is an analytic function in $\Omega^*$. Now we restrict system (\ref{MB1}) to its invariant plane $\Pi$ and rescale the time $t \mapsto \tau$ with $d \tau = \eta \; d t$ to obtain the planar system
\begin{eqnarray}
\frac{dy_1}{d \tau} &=& y_2 + \varepsilon B(\sqrt{2(c-y_2)}, y_1, y_2)  \ , \nonumber \\
\frac{dy_2}{d \tau} &=&  -y_1 - \varepsilon \sqrt{2(c-y_2)} \, A(\sqrt{2(c-y_2)}, y_1, y_2), \label{PlanarSystem-MB}
\end{eqnarray}
which is defined on $\Pi$. Notice that in the particular case in which the perturbation $(A(x), B(x), C(x))$ is  polynomial with $A$ and $B$ even and odd, respectively, in the variable $x_1$, that is having the form $A(x)=  \hat{A}(x_1^2, x_2, x_3)$ and $B(x) = x_1 \hat{B}(x_1^2, x_2, x_3)$ then (\ref{PlanarSystem-MB}) is also a polynomial perturbation of the harmonic oscillator.

\subsection{Example: Euler top}

As a second instance of the reduction procedure consider the Euler equations, which describe the rotation of a rigid body:
\begin{equation}
\label{top}
    \dot{x}_1 = \frac{\mu _2 - \mu _3}{\mu _2 \mu _3}x_2x_3  \:\: , \:\:\:
    \dot{x}_2 = \frac{\mu _3 - \mu _1}{\mu _3 \mu _1}x_3x_1  \:\: , \:\:\:
    \dot{x}_3 = \frac{\mu _1 - \mu _2}{\mu _1 \mu _2}x_1x_2  \:\: . 
\end{equation}
In system (\ref{top}) each variable $x_i$ denotes the $i$th component of angular momentum, and constants $\mu _i$ are the moments of inertia about the coordinate axes, both for $i=1,2,3$. Energy is conserved for this vector field, and actually this is a Poisson system \cite{arn1,olv1} in terms of the following structure matrix:
\[
     {\cal J}(x) = \left( \begin{array}{ccc}
                    0  & -x_3 & x_2  \\
                   x_3 &  0   & -x_1 \\
                   -x_2 & x_1  &  0
              \end{array} \right) \:\: .
\]
Obviously the rank of the structure matrix is 2 everywhere in $\mathbb{R}^3$ except at the origin. The Hamiltonian, which is the total (kinetic) energy, can be written as:
\[
    H(x) = \frac{1}{2} \left( \frac{x_1^2}{\mu _1} + \frac{x_2^2}{\mu _2} +
    \frac{x_3^2}{\mu _3} \right) \ .
\]
Euler top has received a significant attention in the Poisson system framework, for instance see \cite{iyb1} and references therein. From the point of view of the study of periodic solution bifurcations after perturbations of the Euler top, see \cite{Bu-Ga}. 
Excluding the origin, there is one independent Casimir invariant, which can be taken as the square of the angular momentum norm:
\[
    D(x) = x_1^2+x_2^2+x_3^2 \ .
\]
Accordingly, we shall denote the symplectic leaves as $\mathcal{L}_{c^2} \equiv \{ x \in \mathbb{R}^3 \: : \: D(x) = c^2\}$. The Hamiltonian is quasi-harmonic for every pair of variables. For instance, in terms of $x_1$ and $x_2$ we have $H(x)= \varphi_1(x_1,x_2)+\varphi_2(x_1,x_2)+\frac{1}{2\mu_3}D(x)$, where $\varphi_i(x_1,x_2)=\frac{1}{2}\kappa^2_{i3}x_i^2$, for $i=1,2$, and 
\[
\kappa_{i3}=\left( \frac{1}{\mu_i} - \frac{1}{\mu_3} \right)^{1/2} \: .
\]
According to the reduction procedure assumptions, we have $\varphi_i(x_1,x_2) \neq 0$ provided $x_1 \neq 0$ and $x_2 \neq 0$, and in addition we assume without loss of generality $\mu_3 > \mu_1$ and $\mu_3 > \mu_2$. Let us also define the semispheres
\[
\mathcal{L}_{c^2}^{+} :=  \{ (x_1, x_2, x_3) \in \mathcal{L}_{c^2} \ : \ x_3 > 0 \} \ , \ \mathcal{L}_{c^2}^{-} :=  \{ (x_1, x_2, x_3) \in \mathcal{L}_{c^2} \ : \ x_3 < 0 \} \ .
\]
Consider now the most general analytic perturbation in $\mathbb{R}^3 \backslash \{ x_3 = 0 \}$ of the Euler top, leaving invariant the semispheres $\mathcal{L}_{c^2}^{+}$ and $\mathcal{L}_{c^2}^{-}$:
\begin{eqnarray}
\dot{x}_1  &=& \frac{\mu _2 - \mu _3} x_2 x_3 + \varepsilon A(x_1,x_2,x_3) \ , \nonumber \\
\dot{x}_2  &=& \frac{\mu _3 - \mu _1}{\mu _3 \mu _1} x_1 x_3 + \varepsilon B(x_1,x_2,x_3)  
\ , \label{Euler9} \\
\dot{x}_3  &=& \frac{\mu _1 - \mu _2}{\mu _1 \mu _2} x_1 x_2 + \varepsilon C(x_1,x_2,x_3) \ , \nonumber
\end{eqnarray}
where
\begin{eqnarray}
A(x_1,x_2,x_3)  &=& x_3 P(x_1,x_2, D(x_1,x_2,x_3)) \ , \nonumber \\
B(x_1,x_2,x_3)  &=& x_3 Q(x_1,x_2, D(x_1,x_2,x_3))  \ , \nonumber \\
C(x_1,x_2,x_3)  &=& \frac{D(x_1,x_2,x_3)-c^2}{2 x_3}  R(x_1,x_2, D(x_1,x_2,x_3)) \nonumber \\
 & & -x_1 P(x_1,x_2, D(x_1,x_2,x_3)) - x_2 Q(x_1,x_2, D(x_1,x_2,x_3)) \ , \nonumber
\end{eqnarray}
with $P$, $Q$ and $R$ analytic functions everywhere in $\mathbb{R}^3$. We then perform the following diffeomorphic change of variables:
\begin{equation}
\label{mistetas}
(x_1, x_2, x_3) \mapsto (y_1,y_2,y_3) = (\kappa_{13}x_1,\kappa_{23}x_2,D(x_1, x_2, x_3)) \ ,
\end{equation}
defined in $\Omega \equiv \{(x_1,x_2,x_3) \in \mathbb{R}^3 : \: x_1 \neq 0, x_2 \neq 0, x_3 \neq 0 \}$. The perturbed system (\ref{Euler9}) restricted to $\mathcal{L}_{c^2}^+$ adopts the form
\begin{eqnarray}
\dot{y}_1 &=& -\kappa_{13} \kappa_{23} \sqrt{y_3-(y_1/ \kappa_{13})^2-(y_2/ \kappa_{23})^2} \left( \frac{\partial H}{\partial y_2} + \varepsilon P(y_1, y_2, y_3) \right) \ , \nonumber \\
\dot{y}_2 &=& -\kappa_{13} \kappa_{23} \sqrt{y_3-(y_1/ \kappa_{13})^2-(y_2/ \kappa_{23})^2} \left( -\frac{\partial H}{\partial y_1} + \varepsilon Q(y_1, y_2, y_3) \right)  \ , \label{Euler12} \\
\dot{y}_3 &=& \varepsilon (y_3-c^2) R(y_1, y_2, y_3) \ , \nonumber
\end{eqnarray}
with $H(y_1,y_2,y_3) = \frac{1}{2} (y_1^2 + y_2^2)+\frac{1}{2 \mu_3}y_3$. The perturbed system (\ref{Euler9}) restricted to the semispace $x_3 < 0$ is given by (\ref{Euler12}) changing the sign in the right-hand side of $\dot{y}_1$ and $\dot{y}_2$. Then, the restriction of system (\ref{Euler12}) to $\mathcal{L}_{c^2}^+$ is given by the analytic system
\begin{eqnarray*}
\dot{y}_1 &=& -\kappa_{13} \kappa_{23} \sqrt{c^2-(y_1/ \kappa_{13})^2-(y_2/ \kappa_{23})^2} \left( \frac{\partial \mathcal{H}}{\partial y_2} + \varepsilon P(y_1, y_2, c^2) \right) \ , \\
\dot{y}_2 &=& -\kappa_{13} \kappa_{23} \sqrt{c^2-(y_1/ \kappa_{13})^2-(y_2/ \kappa_{23})^2} \left( -\frac{\partial \mathcal{H}}{\partial y_1} + \varepsilon Q(y_1, y_2, c^2) \right)  \ ,
\end{eqnarray*}
where $\mathcal{H}(y_1,y_2)= \frac{1}{2} (y_1^2 + y_2^2)$. Finally, we introduce a time reparametrization $t \mapsto \tau$ of the form $d \tau = \eta \; d t$, with $\eta = -\kappa_{13} \kappa_{23} \sqrt{c^2-(y_1/ \kappa_{13})^2-(y_2/ \kappa_{23})^2}$ which completes the reduction to the form (\ref{PlanarSystem*}) of a perturbed harmonic oscillator.

\section{Perturbations of the harmonic oscillator}

As far as we know, the bifurcation of limit cycles from the period annulus $\mathcal{P} = \mathbb{R}^2 \backslash \{(0,0)\}$ of a harmonic oscillator $\dot{y}_1 = y_2 + \varepsilon P$, $\dot{y}_2 = -y_1 + \varepsilon Q$ was first analyzed in \cite{GGV} for polynomial perturbation fields $(P(y_1,y_2), Q(y_1,y_2))$ of arbitrary degree and whose coefficients are independent of $\varepsilon$. The {\it cyclicity} of $\mathcal{P}$ under perturbations $(P,Q)$ with $|\varepsilon| \ll 1$ is the maximum number of limit cycles bifurcating from the circles that foliates $\mathcal{P}$. A detailed analysis of the homogeneous case for which $P$ and $Q$ are homogeneous polynomials of the same arbitrary degree is given in \cite{GM} where its is shown that the cyclicity of $\mathcal{P}$ is zero.

Later in \cite{I} the cyclicity of $\mathcal{P}$ under arbitrary polynomial perturbation fields $(P(y_1,y_2; \varepsilon), Q(y_1,y_2; \varepsilon))$ is analyzed but now allowing the coefficients to depend analytically on $\varepsilon$, that is $P, Q \in \mathbb{R}\{\varepsilon\}[y_1, y_2]$. In \cite{I} it is derived the global upper bound $[\ell (n-1)/2]$ on the cyclicity of $\mathcal{P}$ where $n = \max \{ \deg(P), \deg(Q) \}$ and $\ell$ is the order of the associated first Melnikov function which is not identically zero. Also in \cite{I} some cases where the above upper bound is sharp are shown.

An interesting question arises if we assume that $P, Q \in \mathbb{R}_m[\varepsilon][y_1, y_2]$, that is, the coefficients of $(P,Q)$ are polynomial functions of $\varepsilon$ having some fixed maximum degree $m$: to find the bifurcation diagram of limit cycles in $\mathcal{P}$ in the parameter space. We consider here the simplest case with respect to the degrees, namely, $(m,n) = (1,2)$. Thus we consider the most general perturbation of a harmonic oscillator like (\ref{PlanarSystem*}) by a quadratic perturbation field $(P,Q)$ whose coefficients are linear functions of the perturbation parameter $\varepsilon$. Moreover, the right hand side can be taken without loss of generality (after a rotation in the phase plane) in the called Bautin form (see \cite{B})
\begin{eqnarray}\label{HO-PS}
\dot{y}_1 &=& -y_2 + \varepsilon \big[ -A_3(\varepsilon) y_1^2 + (2 A_2(\varepsilon) + A_5(\varepsilon)) y_1 y_2 + A_6(\varepsilon) y_2^2 \big], \\
\dot{y}_2 &=& y_1 + \varepsilon \big[ A_2(\varepsilon) y_1^2 + (2 A_3(\varepsilon) + A_4(\varepsilon)) y_1 y_2 -A_2(\varepsilon) y_2^2 \big], \nonumber
\end{eqnarray}
with linear coefficients $A_{i}(\varepsilon) = a_{i0} + a_{i1} \varepsilon$ for $i=2,3, 4, 5, 6$. The resulting perturbation coefficients $a_{ij}$ are collected into the vector parameter $\lambda \in \mathbb{R}^{10}$.

\begin{remark}\label{Q-cent}
{\rm After \cite{B}, it is well known that the origin is a center of family (\ref{HO-PS}) for any $\varepsilon \in \mathbb{R}$ if and only if one of the following four conditions is fulfilled:
\begin{enumerate}
\item[(a)] $A_4(\varepsilon) = A_5(\varepsilon) \equiv 0$;

\item[(b)] $A_3(\varepsilon) - A_6(\varepsilon) \equiv 0$;

\item[(c)] $A_5(\varepsilon)  = A_4(\varepsilon) + 5(A_3(\varepsilon)-A_6(\varepsilon)) = A_3(\varepsilon) A_6(\varepsilon) -2 A_6^2(\varepsilon) -A_2^2(\varepsilon) \equiv 0$;

\item[(d)] $A_2(\varepsilon) = A_5(\varepsilon) \equiv 0$.
\end{enumerate} }
\end{remark}

Introducing polar coordinates $y_1= r \cos\theta$, $y_2= r \sin\theta$, and for $|\varepsilon|$ sufficiently small, any system $\dot{y}_1 = y_2 + \varepsilon P(y_1,y_2; \varepsilon)$, $\dot{y}_2 = -y_1 + \varepsilon Q(y_1,y_2; \varepsilon)$ and in particular system (\ref{HO-PS}) is transformed into the analytic differential equation
\begin{equation}\label{HO2}
\frac{d r}{d \theta} = \mathcal{F}(\theta, r; \lambda, \varepsilon)
\end{equation}
which is defined on the cylinder $\{ (r, \theta) \in (\mathbb{R}^+ \cup \{0\}) \times \mathbb{S}^1 \}$ with $\mathbb{S}^1  = \mathbb{R}/ 2 \pi \mathbb{Z}$ and satisfies $\mathcal{F}(\theta, r; \lambda, 0) \equiv 0$. Therefore, equation (\ref{HO2}) is written in the standard Lagrange form of the averaging theory with period $2 \pi$. The classical tool of averaging allows us to analyze the $2 \pi$-periodic solutions of (\ref{HO2}), see for example the book \cite{SVM} or, for recent advances, the papers \cite{GLM} and \cite{LNT}.

The solution $r(\theta; z, \lambda, \varepsilon)$ of (\ref{HO2}) with initial condition $r(0; z, \lambda, \varepsilon) = z \in \mathbb{R}^+$ admits the convergent power series expansion near $\varepsilon=0$ like $r(\theta; z, \lambda, \varepsilon) = z + \sum_{j \geq 1} r_j(\theta, z, \lambda) \, \varepsilon^j$ where the coefficient functions $r_j$ are real analytic. The function $r(.; z, \lambda, \varepsilon)$ is defined on the interval $[0, 2 \pi]$ provided that $\varepsilon$ is close enough to $0$, hence we can define the {\it displacement map} $d : \mathbb{R}^+ \times \mathbb{R}^{12} \times I \to \mathbb{R}^+$ with $I$ some real interval containing the origin as $d(z, \lambda, \varepsilon) = r(2 \pi; z, \lambda,  \varepsilon) - z$. From this definition we see that the isolated positive zeros $z_0 \in \mathbb{R}^+$ of $d(., \lambda, \varepsilon)$ are just the initial conditions for the $2 \pi$-periodic solutions of (\ref{HO2}), which clearly are in one-to-one correspondence with the limit cycles of system (\ref{HO-PS}) bifurcating from the circle $y_1^2+y_2^2 = z_0^2$ included in the period annulus $\mathcal{P}$ of the unperturbed harmonic oscillator.

In summary, the displacement map $d$ is expressed as the following convergent series expansion
\[
d(z, \lambda, \varepsilon) = \sum_{i \geq 1}  f_i(z; \lambda) \, \varepsilon^i,
\]
and the coefficient functions $f_i(z; \lambda) = r_i(2 \pi, z, \lambda)$ can be computed by a recursive procedure, see \cite{grn} for the general structure. We call $f_i$ the $i$-th {\it averaged function} (also called $i$-th Melnikov function in the literature).

We say that a branch of limit cycles bifurcates from the circle $y_1^2+y_2^2 = z_0^2$ with $z_0 \in \mathbb{R}^+$ if there is a function $z^*(\lambda, \varepsilon)$ (which may be defined only for values of $\varepsilon$ on a half-neighborhood of zero) such that $z^*(\lambda, 0) = z_0$ and $d(z^*(\lambda, \varepsilon), \lambda, \varepsilon) \equiv 0$. It is well known (see \cite{SVM}, for example) that in such a case $z_0$ must be a zero of the function $f_\ell(.; \lambda)$ where $\ell$ is the first subindex such that $f_\ell(z; \lambda) \not\equiv 0$, that is the first non-identically zero averaged function is the $\ell$-th.

\begin{remark}\label{Q-cent-saturation}
{\rm Since the averaged functions $f_i(z; \lambda) = z^{m_j} \sum_{j=0}^{n_j} \xi_{ij}(\lambda) \, z^j \in \mathbb{R}[\lambda][z]$, we can consider the polynomial ideal $\mathcal{I}$ generated by its coefficients $\xi_{ij} \in \mathbb{R}[\lambda]$ in the ring $\mathbb{R}[\lambda]$. We also can consider the ascending chain of ideals
$$
\mathcal{I}_{2} \subseteq \mathcal{I}_{3} \subseteq \cdots \subseteq \mathcal{I}_{k} = \mathcal{I}
$$
where $\mathcal{I}_s = \langle \xi_{ij} : 2 \leq i \leq s \rangle$. Since $\mathcal{I}$ is a Noetherian ring, the above chain stabilizes at, say, the moment $k \in \mathbb{N}$. The former implies that if the parameters $\lambda = \lambda^* \in \mathcal{I}_{k}$, then $d(z, \lambda^*, \varepsilon) \equiv 0$ and the origin becomes a center of (\ref{HO-PS}).  }
\end{remark}

\begin{remark}\label{Braches bounds}
{\rm We summarize here the classical averaging theory applied to the differential equation (\ref{HO2}). Assume that $z_0 \in \mathbb{R}^+$ is a zero of $f_\ell(.; \lambda^*)$, the first non identically zero averaged function and let $N$ be the number of isolated branches of $2 \pi$-periodic solutions of (\ref{HO2}) with parameters $\lambda = \lambda^*$ bifurcating from $z_0$ for $|\varepsilon| \ll 1$. Then the following statements
hold:
\begin{itemize}
\item[(i)] If $z_0$ is simple then $N=1$.

\item[(ii)] If $z_0$ is multiple of multiplicity $\bar k$, then $N \leq \bar k$.
\end{itemize}
Notice that (i) is a simple consequence of the Implicit Function Theorem while for (ii) it is required the Weierstrass Preparation Theorem. }
\end{remark}

\begin{theorem}\label{Teo-QPert}
Let us consider the perturbed harmonic oscillator given by family (\ref{HO-PS}) and the following set of polynomials in their parameters $\lambda \in \mathbb{R}^{10}$:
\begin{eqnarray*}
\xi_{20}(\lambda) &=& a_{50} (a_{30} - a_{60}), \\
\hat{\xi}_{30}(\lambda) &=& a_{31} a_{50} + a_{30} a_{51} - a_{51} a_{60} - a_{50} a_{61},  \\
\hat{\xi}_{40}(\lambda) &=& a_{51} (a_{31} - a_{61}), \\
\hat{\xi}_{42}(\lambda) &=& -a_{20} a_{40} (5 a_{30} + a_{40} - 5 a_{60}) (a_{30} - a_{60}), \\
\hat{\xi}_{51}(\lambda) &=& 5 a_{21} a_{30}^2 a_{40} + 10 a_{20} a_{30} a_{31} a_{40} + a_{21} a_{30} a_{40}^2 +
 a_{20} a_{31} a_{40}^2 + 5 a_{20} a_{30}^2 a_{41} + \\
 & & 2 a_{20} a_{30} a_{40} a_{41} -
 10 a_{21} a_{30} a_{40} a_{60} - 10 a_{20} a_{31} a_{40} a_{60} - a_{21} a_{40}^2 a_{60} - \\
 & & 10 a_{20} a_{30} a_{41} a_{60} - 2 a_{20} a_{40} a_{41} a_{60} + 5 a_{21} a_{40} a_{60}^2 +
 5 a_{20} a_{41} a_{60}^2 - \\
 & & 10 a_{20} a_{30} a_{40} a_{61} - a_{20} a_{40}^2 a_{61} +
 10 a_{20} a_{40} a_{60} a_{61}, \\
\hat{\xi}_{61}(\lambda) &=& -10 a_{21} a_{30} a_{31} a_{40} - 5 a_{20} a_{31}^2 a_{40} - a_{21} a_{31} a_{40}^2 -
 5 a_{21} a_{30}^2 a_{41} - 10 a_{20} a_{30} a_{31} a_{41} - \\
 & & 2 a_{21} a_{30} a_{40} a_{41} -
 2 a_{20} a_{31} a_{40} a_{41} - a_{20} a_{30} a_{41}^2 + 10 a_{21} a_{31} a_{40} a_{60} + \\
 & & 10 a_{21} a_{30} a_{41} a_{60} + 10 a_{20} a_{31} a_{41} a_{60} + 2 a_{21} a_{40} a_{41} a_{60} +
 a_{20} a_{41}^2 a_{60} - \\
 & & 5 a_{21} a_{41} a_{60}^2 + 10 a_{21} a_{30} a_{40} a_{61} +
 10 a_{20} a_{31} a_{40} a_{61} + a_{21} a_{40}^2 a_{61} + \\
 & &  10 a_{20} a_{30} a_{41} a_{61} +
 2 a_{20} a_{40} a_{41} a_{61} - 10 a_{21} a_{40} a_{60} a_{61} - 10 a_{20} a_{41} a_{60} a_{61} - \\
 & & 5 a_{20} a_{40} a_{61}^2, \\
\hat{\xi}_{63}(\lambda) &=& a_{20} a_{40}^2 (a_{30} - a_{60}) (5 a_{20}^2 + a_{40} a_{60} + 5 a_{60}^2).
\end{eqnarray*}
Let $N(\lambda)$ be the number of limit cycles that bifurcate from its period annulus $\mathcal{P} = \mathbb{R}^2 \backslash \{ (0,0) \}$ as the perturbation parameter $\varepsilon$ slightly varies from zero. Then the following holds:
\begin{itemize}
\item[(i)] If $\xi_{20} \neq 0$ then $N=0$;

\item[(ii)] If $\xi_{20} = 0$ and $\hat{\xi}_{30} \neq 0$ then $N=0$;

\item[(iii)] If $\xi_{20} = \hat{\xi}_{30} = \hat{\xi}_{42} = 0$ then $N=0$;

\item[(iv)] If $\xi_{20} = \hat{\xi}_{30} = 0$ but $\hat{\xi}_{42} \neq 0$ then, defining $s_1 = \hat{\xi}_{40} / \hat{\xi}_{42}$, we have that $N=1$ or $N=0$ according to wether $s_1 <0$ or $s_1 \geq 0$, respectively;

\item[(v)] If $\xi_{20}= \hat{\xi}_{30} = \hat{\xi}_{40} = \hat{\xi}_{42} = 0$ and $\hat{\xi}_{51} \neq 0$ then $N=0$;

\item[(vi)] If $\xi_{20} = \hat{\xi}_{30} = \hat{\xi}_{40} = \hat{\xi}_{42} = \hat{\xi}_{51} = 0$ but $\hat\xi_{63} \neq $ then, defining $s_2 = \hat\xi_{61}/\hat\xi_{63}$, we have that $N=1$ or $N=0$ according to whether $s_2 <0$ or $s_2 \geq 0$, respectively.
\end{itemize}
\end{theorem}
{\it Proof}. Straightforward computations produce the following averaged functions for system (\ref{HO-PS}):
\begin{eqnarray*}
f_1(z; \lambda) &\equiv& 0, \\
f_2(z; \lambda) &=& z^3 \, \xi_{20}(\lambda), \\
f_3(z; \lambda) &=& z^3 \, [\xi_{30}(\lambda) + z \xi_{31}(\lambda)], \\
f_4(z; \lambda) &=& z^3 \, [\xi_{40}(\lambda) + z \xi_{41}(\lambda) + z^2 \xi_{42}(\lambda)], \\
f_5(z; \lambda) &=& z^4 \, [\xi_{50}(\lambda) + z \xi_{51}(\lambda) + z^2 \xi_{52}(\lambda)] ,\\
f_6(z; \lambda) &=& z^4 \, [\xi_{60}(\lambda) + z \xi_{61}(\lambda) + z^2 \xi_{62}(\lambda) + z^3 \xi_{63}(\lambda)],
\end{eqnarray*}
where $\xi_{i j} \in \mathbb{R}[\lambda]$ are the polynomials in the parameters of family (\ref{HO-PS}). In what follows we shall denote by $\hat{\xi}_{ij}$ the remainder of $\xi_{ij}$ upon division by a Gr\"obner basis of the ideal generated by all the $\xi_{ks}$ with $k < i$ in the polynomial ring $\mathbb{R}[\lambda]$. This remainder can be computed, for instance, with the functions \texttt{PolynomialReduce} and \texttt{GroebnerBasis} of the computer algebra system Mathematica$^{\tiny \copyright}$. Another option is the use of \texttt{reduce} with the software Singular$^{\tiny \copyright}$. The non-identically zero polynomials $\hat{\xi}_{ij} \in \mathbb{R}[\lambda]$ are listed in the statement of the theorem. After such reduction, we will consider the polynomials:
\begin{eqnarray*}
f_2(z; \lambda) &=& z^3 \, \xi_{20}(\lambda), \\
\hat{f}_3(z; \lambda) &=& \hat{\xi}_{30}(\lambda) z^3, \\
\hat{f}_4(z; \lambda) &=& z^3 \, [\hat{\xi}_{40}(\lambda) +  z^2 \hat{\xi}_{42}(\lambda)], \\
\hat{f}_5(z; \lambda) &=& \hat{\xi}_{51}(\lambda) \, z^5,\\
\hat{f}_6(z; \lambda) &=& z^5 \, [\hat{\xi}_{61}(\lambda) + z^2 \hat{\xi}_{63}(\lambda)] \: .
\end{eqnarray*}
From the expression of $f_2(z; \lambda)$ and $\hat{f}_3(z; \lambda)$ we deduce statements (i) and (ii) respectively while from the expression of $\hat{f}_4(z; \lambda)$ we obtain (iii) and (iv). Next (v) and (vi) are obtained from the expressions of $\hat{f}_5(z; \lambda)$ and $\hat{f}_6(z; \lambda)$. $\Box$
\newline

Notice that the complete limit cycle bifurcation diagram of $\mathcal{P}$ in the parameter space $\mathbb{R}^{10}$ for family (\ref{HO-PS}) when $\xi_{20} = \hat{\xi}_{30} = \hat{\xi}_{40} = \hat{\xi}_{42} = \hat{\xi}_{51} = \hat\xi_{61} = \hat\xi_{63} = 0$ (that is for parameters $\lambda = \lambda^*$ lying in the real variety associated with $\mathcal{I}_{6}$) is not presented. Unfortunately the massive computations to obtain $f_7(z; \lambda)$, hence $\hat{f}_7(z; \lambda)$, in the proof of Theorem \ref{Teo-QPert} do not seem to be possible in our computer. In other words, for family (\ref{HO-PS}) we are unable to get the ideal stabilization explained in Remark \ref{Q-cent-saturation}. The reason is that we can check that $\mathcal{I}_{6} \neq \mathcal{I}$ because there are parameters in $\mathcal{I}_{6}$ for which the origin is not a center of (\ref{HO-PS}) as can be easily seen by using Remark \ref{Q-cent}. Anyway, the  bifurcation diagram can be made complete with a further case-by-case explicit analysis of the 10 subcases that arise after the vanishing of the factors in the expressions of $\xi_{20}$, $\xi_{40}$ and $\xi_{42}$ which are the simpler ones.
\newline

We remark on the other hand that in all the cases exposed in Theorem \ref{Teo-QPert} we have obtained simple zeroes of the corresponding averaged function $f_\ell(.; \lambda)$. In order to compute the actual value (not only its upper bound as in part (ii) of Remark \ref{Braches bounds}) of the number of branches bifurcating from a multiple zero $z_0$ of $f_\ell(.; \lambda)$ several methods can be employed. Among them branching theory and singularity theory applied to the reduced displacement map $\delta(z, \lambda, \varepsilon) = f_\ell(z; \lambda) + \sum_{i \geq 1} f_{\ell + i}(z; \lambda) \varepsilon^i$ are worth mentioning. Branching theory uses the Newton's diagram of $\delta$ (see \cite{VT}) to analyze the local structure of the zeroes of $\delta$ near $(z, \varepsilon) =(z_0, 0)$. The approach of singularity theory of smooth functions (see for example \cite{GS}) is completely different: the goal is to find when $\lambda=\lambda^*$ a normal form $\hat\delta(z, \varepsilon)$ of $\delta(z, \varepsilon)$ such that $U(z, \varepsilon) \, \delta(Z(z, \varepsilon), \Lambda(\varepsilon)) = \hat\delta(z, \varepsilon)$ where $(z, \varepsilon) \mapsto (Z(z, \varepsilon), \Lambda(\varepsilon))$ is a local diffeomorphism of $\mathbb{R}^2$ mapping the origin to $(z_0, 0)$ and preserving orientation whereas $U(z, \varepsilon) > 0$. A different approach dealing with the degenerate case for which $z_0$ is a multiple zero of $f_\ell(.; \lambda)$ and $f_k(z_0; \lambda) = 0$ for any $k \in \mathbb{N}$ can be found in  \cite{GLM}.
\newline

In the next example, the analysis of multiple zeroes of $f_\ell(.; \lambda)$ is needed.

\begin{proposition}\label{prop-3Pert}
Let us consider the perturbed harmonic oscillator given by system $\dot{y}_1 = y_2 + \varepsilon P_3(y_1,y_2; \varepsilon)$, $\dot{y}_2 = -y_1 + \varepsilon Q_3(y_1,y_2; \varepsilon)$ with the cubic perturbation
\begin{eqnarray*}
P_3(y_1,y_2; \varepsilon) &=& \left( \frac{289}{2} + \frac{18719 \, \varepsilon}{884736 \beta} \right) x^3- \frac{1}{4} \beta x^2 y - \frac{867}{2} x y^2 + \frac{\beta}{12} y^3, \\
Q_3(y_1,y_2; \varepsilon) &=& - \frac{1}{768} x y + \varepsilon y^2  - \frac{861}{2} x^2 y + \left( \frac{287}{2} - \frac{18719 \, \varepsilon}{884736 \beta} \right) y^3,
\end{eqnarray*}
and $\beta = \sqrt{145}$. Then limit cycles on the period annulus $\mathcal{P}$ only can bifurcate from the circle $x^2+y^2 = 1/2$. Moreover, exactly either two or none limit cycles bifurcate according to whether $\varepsilon > 0$ or $\varepsilon < 0$, respectively.
\end{proposition}
{\it Proof}. Straightforward computations produce the following averaged functions for system (\ref{HO-PS}):
\begin{eqnarray*}
f_1(z) &=& f_2(z) \equiv 0, \\
f_3(z) &=& z^3 \, (-1 + 2 z^2)^2, \\
f_4(z) &=& z^5 \, (8210368799 - 21687552313344 z^2 + 295572602880 z^4).
\end{eqnarray*}
Therefore, the reduced displacement map $\delta(z, \varepsilon) = d(z, \varepsilon) / \varepsilon^3$ has the form $\delta(z, \varepsilon) = f_3(z) + f_4(z) \, \varepsilon + \mathcal{O}(\varepsilon^2)$ where $z_0 = \sqrt{2}/2 \in \mathbb{R}^+$ is a multiple zero of $f_3$ of multiplicity $\bar{k} = 2$. We know then that at most 2 limit cycles can bifurcate from the circle $x^2+y^2 = z_0^2$. The following analysis will show that actually this bound is sharp. Indeed, since $f_4(z_0) \neq 0$, using singularity theory of smooth maps (see \cite{GS}), we deduce that $\delta$ is strongly equivalent to the normal form $\tilde\delta(z, \varepsilon) = \delta_1 z^{2} + \delta_2 \varepsilon$ where $\delta_j$ are $\pm 1$ according to the signs
$$
\delta_1 = {\rm sgn} \left( \frac{d^2 f_3}{d z^2}(z_0) \right) \neq 0, \ \ \ \delta_2 = {\rm sgn}\left( f_{4}(z_0) \right) \neq 0.
$$
We recall here that $\tilde\delta(z,  \varepsilon)$ and $\delta(z, \varepsilon)$ are strongly equivalent if they are related by $U(z, \varepsilon) \, \delta(Z(z, \varepsilon), \varepsilon) = \tilde\delta(z, \varepsilon)$ where $z \mapsto Z(z, \varepsilon)$ is a local diffeomorphism  of $\mathbb{R}$ mapping the origin to $z_0$ and preserving orientation, and $U(z, \varepsilon)$ is a positive function. Notice that if $N_\delta(\varepsilon)$ denotes the number of local zeros of $\delta(., \varepsilon)$ near $z_0$ and $N_{\tilde\delta}(\varepsilon)$ the number of local zeros of $\tilde\delta(.,  \varepsilon)$ near $0$ then we arrive at the important consequence for our purpose that $N_\delta(\varepsilon) = N_{\tilde\delta}(\varepsilon)$.

In our case $\delta_1 = 1$ and $\delta_2 = -1$ so that $\tilde\delta(., \varepsilon)$ has exactly two zeros $z^*_{\pm}(\varepsilon) = \pm \sqrt{\varepsilon}$ which only appear when $\varepsilon > 0$ so that $z^*_{\pm} \in \mathbb{R}$. Therefore, going back we conclude that exactly two limit cycles bifurcate from the circle $x^2+y^2 = z_0^2$ when $\varepsilon > 0$ and no limit cycle bifurcation occurs with the contrary sign of $\varepsilon$. $\Box$

\mbox{}

\mbox{}

\noindent {\bf Acknowledgments.}

\noindent
Both authors would like to acknowledge partial support from Ministerio de Econom\'{\i}a, Industria y Competitividad for grant MTM2017-84383-P. In addition, I.A.G. acknowledges AGAUR (Generalitat de Catalunya) grant number 2017SGR-1276. B.H.-B. acknowledges Ministerio de Econom\'{\i}a y Competitividad for grant MTM2016-80276-P as well as financial support from Universidad Rey Juan Carlos-Banco de Santander (Excellence Group QUINANOAP, grant 30VCPIGI14).

\pagebreak

\end{document}